\documentclass[lettersize,journal]{IEEEtran}

\usepackage{cite}
\usepackage[colorlinks = true, linkcolor=red, citecolor=red,urlcolor=red]{hyperref}
\usepackage{amsmath,amssymb,amsfonts}
\usepackage{algorithmic}
\usepackage{graphicx}
\usepackage{textcomp}

\def\BibTeX{{\rm B\kern-.05em{\sc i\kern-.025em b}\kern-.08em
		T\kern-.1667em\lower.7ex\hbox{E}\kern-.125emX}}

\graphicspath{{./figures/}}

\begin{document}

\title{How Do Socio-Demographic Patterns\\ Define Digital Privacy Divide?}

\author{\uppercase{Hamoud~Alhazmi}, \uppercase{Ahmed~Imran}, and \uppercase{Mohammad~Abu~Alsheikh}

\thanks{This work was supported in part by the Australian Research Council (ARC) under Grant DE200100863. The data collection in this project has been approved by the Human Research Ethics Committee at the University of Canberra under application 4522 ``Privacy Coupling: When Your Personal Devices Betray You''.}
}

\markboth{Published in IEEE Access, DOI: 10.1109/ACCESS.2022.3144436}%
{DOI: 10.1109/ACCESS.2022.3144436}

\maketitle

\begin{abstract}
Digital privacy has become an essential component of information and communications technology (ICT) systems. There are many existing methods for digital privacy protection, including network security, cryptography, and access control. However, there is still a gap in the digital privacy protection levels available for users. This paper studies the digital privacy divide~(DPD) problem in ICT systems. First, we introduce an online DPD study for understanding the DPD problem by collecting responses from 776~ICT users using crowdsourcing task assignments. Second, we propose a factor analysis-based statistical method for generating the DPD index from a set of observable DPD question variables. In particular, the DPD index provides one scaled measure for the DPD gap by exploring the dimensionality of the eight questions in the DPD survey. Third, we introduce a DPD proportional odds model for analyzing the relationship between the DPD status and the socio-demographic patterns of the users. Our results show that the DPD survey meets the internal consistency reliability with rigorous statistical measures, e.g.,~Cronbach’s $\alpha=0.92$.  Furthermore, the DPD index is shown to capture the underlying communality of all DPD variables. Finally, the DPD proportional odds model indicates a strong statistical correlation between the DPD status and the age groups of the ICT users. For example, we find that young users (15-32 years) are generally more concerned about their digital privacy than senior ones (33 years and over).

\end{abstract}


\section{Introduction}\label{sec:intro}

\IEEEPARstart{R}ecent years have witnessed much progress in defining digital privacy as a functional requirement in information and communications technology (ICT) systems~\cite{sahi2017privacy,xu2014information,nguyen2021security,gupta2020security}. For the first time in human history, digital privacy is well-defined in regulations and policies, including the general data protection regulation~(GDPR)~\cite{gdpr2016general}. Digital privacy can now be measured using statistical tools, e.g.,~differential privacy~\cite{dwork2014algorithmic}. Previous works~\cite{sahi2017privacy,xu2014information,nguyen2021security,gupta2020security} have developed privacy-preserving algorithms that protect the digital privacy of individuals. The literature reflects extensive efforts and attention in digital privacy from the research communities, industries, and governments. Nevertheless, is there still a gap in the levels of digital privacy provided to individuals?

One significant problem of ICT systems is their digital divide and inequality~\cite{pick2015global,lai2021revisiting,estacio2019digital,chaoub20216g,reddick2020determinants,mathrani2020digital}. In particular, ICT users receive various service levels and access qualities based on their socio-demographic patterns, e.g.,~age, gender, geographical location, occupation, and education. This paper shows that digital privacy is a recent form of the digital divide in ICT systems. In particular, the digital privacy divide~(DPD) describes the various levels of digital privacy protection provided to users based on their socio-demographic patterns. This paper provides an in-depth statistical analysis of the effects of  socio-demographic patterns on the DPD gap, which is a critical initial step for addressing the DPD problem and privacy protection inequalities.

We conducted an online survey study between May and October 2021 on how people perceive the DPD problem in their countries of residence. We collected responses from 776~ICT users. The study was created using Qualtrics survey software~\cite{qualtrics2016survey}, and the ICT users were mainly recruited using manual referrals and Amazon Mechanical Turk (MTurk)~\cite{mturk2021amazon} for crowdsourcing task assignments. Previous work shows that crowdsourcing in survey research accurately reflects the general population~\cite{redmiles2019well,behrend2011viability}. Then, we applied rigorous statistical analysis to the collected DPD data. First, we show that the DPD survey meets the requirements for internal consistency reliability, e.g.,~Cronbach’s $\alpha=0.92$ and McDonald’s $\omega=0.94$. Second, our DPD data visualization shows that the users' geolocations cover most parts of Bangladesh, Germany, India, and the United States. Third, the distribution of responses to the DPD questions indicates a similar response pattern among the users in Bangladesh and Germany and those in India and the United States.

We describe how to create the DPD index from a set of eight question variables\footnote{For the rest of this paper, we use ``DPD question variables'' and ``DPD variables'' interchangeably.}. The DPD index is a latent construct, which enables the study of the relationship between the DPD problem and the socio-demographic patterns of users. Our results show that the DPD status, i.e., DPD class, can be defined based on the DPD index. Furthermore, we propose a DPD proportional odds model for analyzing the statistical relationship between the DPD problem and the socio-demographic patterns of the individuals. For example, our results show that young users (15-32 years) are generally more concerned about their digital privacy compared to the more senior ones (33 years and over).

%
%
%
%
%
%

\subsection{Paper organization}

The rest of this paper is organized as follows. Section~\ref{sec:sec_1} presents related works. Section~\ref{sec:sec_2} provides an introduction to privacy and data protection in the digital age, discusses the DPD problem, and introduces the online survey study. Section~\ref{sec:sec_3} introduces the DPD statistical analysis of the DPD index generation and DPD proportional odds model. Then, numerical results are given in Section~\ref{sec:sec_4}. Section~\ref{sec:roadmap} presents recommendations for closing the DPD gap. Finally, conclusions and future works are highlighted in Section~\ref{sec:conclusions}.

\section{Related Works}\label{sec:sec_1}

Related works fall into three areas. We first review the digital divide problem in ICT systems. Then, we discuss digital privacy. Finally, we review related works in survey research with crowdsourcing.

\subsection{Digital divide}

The digital divide problem broadly refers to the uneven distribution, access, and usage of ICT, resulting in opportunity gaps between individuals. Recent years have witnessed much interest in the digital divide, e.g.,~in education~\cite{lai2021revisiting}, digital health~\cite{estacio2019digital}, and the COVID-19 pandemic impact on rural communities~\cite{mueller2021impacts}. Chaoub~\textit{et al.}~\cite{chaoub20216g} discussed rural wireless connectivity and suggested solutions to narrow the digital divide gap for people living in remote areas, including affordability, accessibility, spectrum, power, and maintenance solutions. Reddick~\textit{et al.}~\cite{reddick2020determinants} conducted a survey to study affordability and broadband access of people living in San Antonio, the United States. They showed that the digital divide is not limited to only regional locations but can exist within metropolitan cities.

There are various studies conducted during the COVID-19 pandemic. For example, Mathrani~\textit{et al.}~\cite{mathrani2020digital} discussed the digital gender divide in India and conducted a survey to determine some of the learning challenges that female students encounter during the COVID-19 pandemic. They found that the digital divide gap exists among users who live in metropolitan, semi-metropolitan, and rural areas. Also, the surveyed users tend to agree that e-learning has affected their productivity and interaction with face-to-face discussions.

\subsection{Digital privacy}

Protecting the digital privacy of individuals is a critical requirement in any modern system. Winegar~\textit{et al.}~\cite{winegar2019much} concluded that more than 70\% of users are concerned about their digital privacy. More concerning, some users do not know that their data is being collected in ICT systems. Jacobson~\textit{et al.}~\cite{jacobson2020social} conducted a survey on the privacy concerns of social media users. They found that users are worried about how their social media data is used in targeted advertising.

Redmiles~\textit{et al.}~\cite{redmiles2017digital} conducted a telephone survey in the United States to study the correlation between the socio-economic status of users and their self-reported data breaches and privacy incidents. They found that advice resources are strongly related to the expected privacy incidents. Also, the authors reported that people from different socio-economic backgrounds might have different views about privacy-related problems.

Digital privacy has several legal and ethical aspects. Minin~\textit{et al.}~\cite{di2021address} investigated the legal basis for personal data processing and using social media data in studying the ecology of human-nature interactions. They addressed the digital privacy problem based on the GDPR privacy rights and suggested applying data anonymization and secure data management to reduce the risk of data exposure. Solove~\cite{solove2021myth} argued that the self-management of privacy by users is not practical in digital privacy protection, e.g.,~users are widely requested to provide personal data to access an online service in a ``take it or leave it'' setup. Instead, privacy laws and regulations, such as the GDPR, should be enforced on institutions to manage data collection, transmission, and processing.

\begin{figure*}
	\begin{centering}
		\includegraphics[width=0.9\textwidth,trim=0cm 1cm 1cm 1cm]{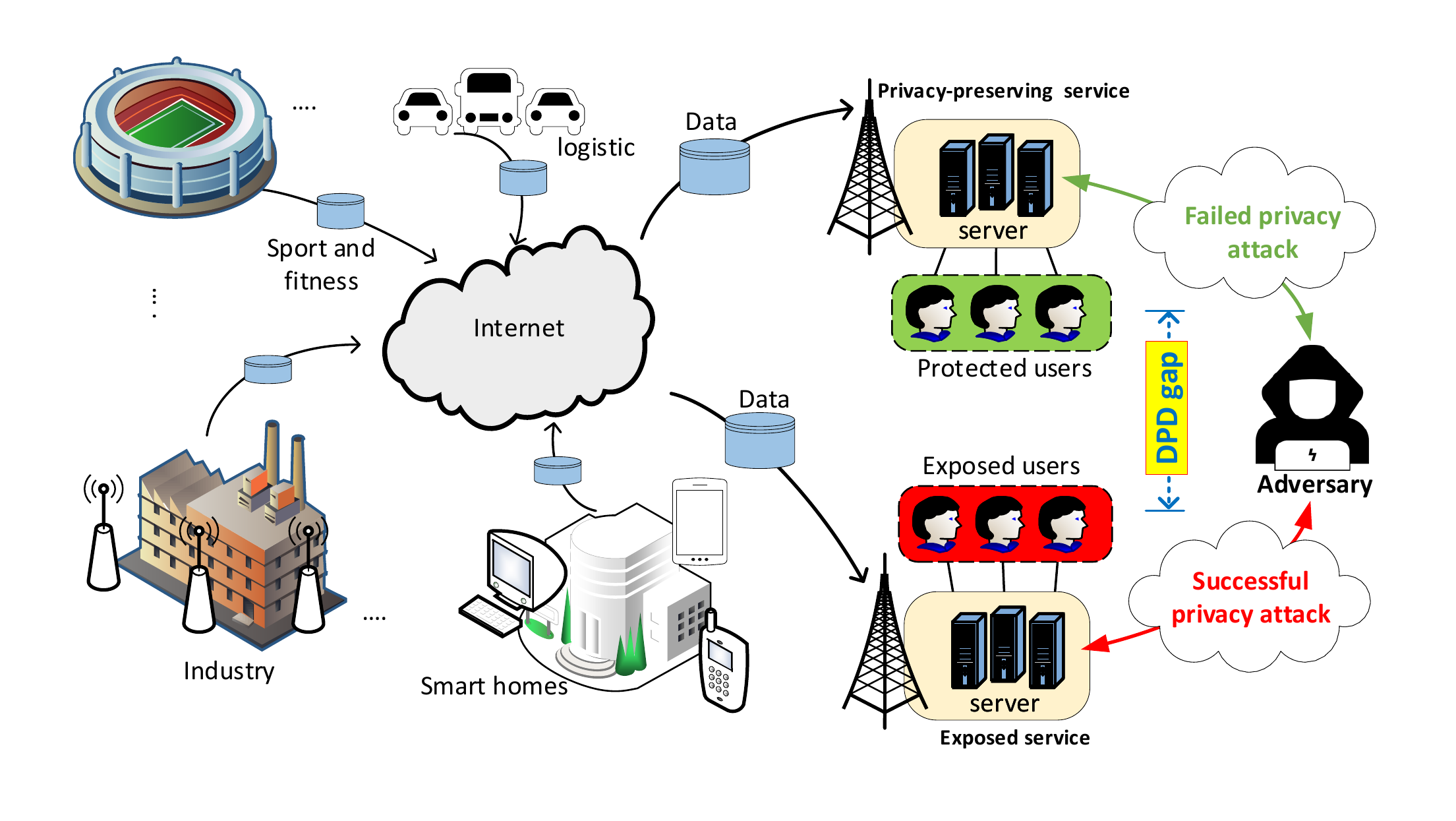}
		\par\end{centering}
	
	\caption{The DPD gap exists between protected and exposed users.\label{fig:system_model}}
\end{figure*}

\subsection{Crowdsourcing in survey research}

Well-designed surveys are an efficient method for collecting responses~\cite{saris2014design}. There are various methods for creating effective online surveys. For example, Story and Tait~\cite{story2019survey} suggested applying reliability measures, explaining the importance of the study to increase the response rate, and assuring that the collected responses are confidentially kept.

Using crowdsourcing, e.g., Amazon MTurk, for completing survey questionnaires is a well-studied area~\cite{redmiles2019well,yigzaw2016secure,yigzaw2016secure}. Redmiles~\textit{et al.}~\cite{redmiles2019well} showed that using MTurk to recruit users for online privacy and security survey is robust and generalizes to the general population. Yigzaw~\textit{et al.}~\cite{yigzaw2016secure} showed that privacy-preserving statistical analysis could be performed on crowdsourcing surveys using secure multi-party computation. Yin~\textit{et al.}~\cite{yin2019privacy} proposed a task recommendation that enables assigning MTurk crowdsourcing tasks to the most fitting users while protecting their digital privacy.

This paper is fundamentally different from all previous works that studied the digital divide and privacy. Our work is the first to address the relationship between the DPD problem and the socio-demographic patterns of ICT users. We used an online survey and recruited 776~users with MTurk crowdsourcing and email invitations. MTurk responses generalize to the general population~\cite{redmiles2019well,behrend2011viability}. Moreover, our DPD survey meets the internal consistency reliability with rigorous statistical measures, e.g.,~Cronbach’s $\alpha=0.92$ and McDonald’s $\omega=0.94$. We conclude this paper with a roadmap discussion on closing the DPD gap among ICT users.

\section{Problem formulation and research methodology}\label{sec:sec_2}
	
This section introduces the reader to the DPD problem and the conducted survey research. First, we provide an introduction to privacy and data protection in the digital age. Second, we describe the DPD problem as a form of digital inequality. Third, we present the procedure of defining the DPD problem using observable questions in the survey study.

\subsection{Primer}
\subsubsection{Privacy and data protection in the digital age}

Modern ICT systems collect massive amounts of data with various ubiquitous and pervasive sensing technologies, such as the Internet of things and crowdsensing~\cite{nguyen20216g}. The plethora of collected data raises genuine concerns about the privacy violation of ICT users. Therefore, \textit{data protection regulations} are essential for dictating how, when, why, and what data is collected. The GDPR~\cite{gdpr2016general} is a fundamental privacy regulation that governs the data collection from residents of the European Union. The GDPR defines \textit{personal data} in Chapter~1, Article~4 as ``any information relating to an identified or identifiable natural person.'' For example, personal data includes browsing cookies, biometric records, and email addresses of users. \textit{Digital privacy} is a concept that describes the right to control how any personal data about users is collected, transmitted, stored, and processed.

\subsubsection{GDPR privacy rights}
The GDPR~\cite{gdpr2016general} defines eight privacy rights for ICT users.
\begin{itemize}
	\item \textit{Right to be informed}: The users must know who collects and obtains their data.
	\item \textit{Right of access}: The users have the right to obtain copies of their data.
	\item \textit{Right to rectification}: The users have the right to request correcting inaccurate records of their data.
	\item \textit{Right to erasure}: The users have the right to be forgotten by deleting their data and preventing future data collection without a new consent.
	\item \textit{Right to restrict processing}: The users have the right to restrict the processing of their data.
	\item \textit{Right to data portability}: The users have the right to transfer their data to selected recipients.
	\item \textit{Right to object}: The users have the right to grant and withdraw consent on processing and collecting their data.
	\item \textit{Rights about automated decision-making and profiling}: The users have the right to opt-out from using their data in automated systems, including machine learning and artificial intelligence~(AI).
\end{itemize}

\subsection{Digital privacy divide~(DPD)}

\textit{Digital privacy divide~(DPD)} is a concept utilized in this work to describe the gap in digital privacy protection between ICT users who are protected and those who are exposed to privacy attacks. Figure~\ref{fig:system_model} depicts the DPD problem in ICT systems. The DPD problem exists when the ICT users receive distinct levels of digital privacy protection. ICT systems enable linking most modern infrastructures, including transportation, industries, and smart homes. ICT systems can be classified into \textit{privacy-preserving} and \textit{exposed} systems in terms of digital privacy protection. Privacy-preserving ICT systems apply rigid privacy tools~\cite{sahi2017privacy,xu2014information,nguyen2021security,gupta2020security}, mitigating the risk of data exposure. Exposed ICT systems do not include well-defined privacy policies, and privacy tools are not properly implemented. Accordingly, data exposure is more likely to occur in exposed systems, where an adversary could obtain private data about exposed users.

\subsubsection{How does the DPD gap influence our digital life?}

The DPD problem produces severe psychological, financial, and social impacts on the exposed users. A{\"\i}meur and Sch{\H{o}}nfeld~\cite{aimeur2011ultimate} discussed identity theft, which can occur due to privacy breaches. They presented several crimes related to identity theft, including financial losses, medical insurance frauds, loan and banking frauds, and criminal impersonation.

\subsubsection{How does the DPD gap connect to other forms of digital inequalities?}

Digital inequalities retain various forms, including physical Internet access and digital literacy. Nevertheless, remarkable progress has been made in recent years to close the gap in physical Internet access and digital literacy. For example, a recent report by Cisco Systems~\cite{cisco2020cisco} estimates that there will be 5.3 billion active Internet users (66\% of the world's population), 5.7 billion mobile subscribers (71\% of the world's population), and 29.3 billion networked devices (3.6 times the world's population) by 2023. Given this rapid increase in accessing online services, more additional users will be affected by the DPD problem over time.

This paper analyzes the DPD problem and its correlation to the socio-demographic patterns of the users. Next, we introduce the design of the online survey study.

\subsection{DPD survey study}

\subsubsection{Choice of operationalization}
Operationalization is the process of representing concepts using observable and measurable questions~\cite{saris2014design}. We developed the DPD survey based on the digital privacy rights in the GDPR~\cite{gdpr2016general}. In particular, the DPD survey includes the following eight questions: 

\begin{itemize}
	\item \textit{Question~1}: I receive clear information on how my government collects my personal data, including who is accessing and processing the data and the data collection purposes.
	\item \textit{Question~2}: I can access copies of my personal data, which my government has collected.
	\item \textit{Q3}: I can transfer my personal data, which my government has collected, to third-party recipients, e.g., organizations, of my choice.
	\item \textit{Question~4}: I can correct my personal data, which my government has collected, when it contains inaccurate, invalid, or misleading data.
	\item \textit{Question~5}: I can request deleting specific records of my personal data, which has been collected by my government when the data is no longer needed for the original purpose.
	\item \textit{Question~6}: I have the option and control to restrict the processing of specific categories of my personal data, which my government has already collected.
	\item \textit{Question~7}: I have the control and ability to grant or withdraw consent on collecting and processing my personal data by my government at any time.
	\item \textit{Question~8}: I have the option and control to opt-out from using my personal data, which my government has collected in making decisions and profiling, based solely on automated processing.
\end{itemize}

Questions~1-8 measure all aspects of the GDPR privacy rights. Utilizing the least possible number of questions in survey research is crucial for improving the response rate and study integrity~\cite{story2019survey}. The ICT users give their responses in a Likert scale~\cite{saris2014design} of five agreement levels (strongly agree, somewhat agree, neither agree nor disagree, somewhat disagree, and strongly disagree). Furthermore, we requested the ICT users to provide responses about their socio-demographic patterns, including their age, gender, ethnicity, highest levels of education,  occupation, and country of residency. The survey responses are presented and analyzed in Section~\ref{sec:sec_4}.

\subsubsection{crowdsourcing for survey research}

We created the survey study using Qualtrics survey software~\cite{qualtrics2016survey}, then we recruited ICT users using Amazon Mechanical Turk (MTurk)~\cite{mturk2021amazon} and manual referrals. The use of survey research with crowdsourcing is a proven method, and the responses generalize to the general population~\cite{redmiles2019well,behrend2011viability}. We collected 776~responses from ICT users residing in Bangladesh, Germany, India, and the United States. We could not recruit many ICT users from Bangladesh with MTurk; therefore, most of the responses from Bangladesh were collected using manual referrals through email invitations.

\section{DPD statistical analysis}\label{sec:sec_3}

This section presents an in-depth statistical analysis of the DPD problem. First, we present statistical methods for defining the DPD index using the observable variables, i.e.,~Questions~1-8. Second, we present a proportional odds model for determining the probability of a DPD status, given the socio-demographic patterns of the ICT users.

\subsection{DPD index generation}

Next, we describe the generation of the DPD index, which provides a single measure of the underlying DPD gap. The DPD index is a latent variable, i.e.,~the DPD index is not directly observable through responses, which articulates the underlying gap in the privacy protection of the ICT users. The DPD questions presented in Section~\ref{sec:sec_2} are unidimensional, i.e., they measure the DPD gap as a single construct.

Principal component analysis (PCA) can be used to reduce the dimensionality of data. Xu~\textit{et al.}~\cite{xu2015pca} proposed PCA-guided clustering for finding the optimal solution of a clustering problem in the PCA subspace. The \textit{DPD status} can be generated by applying the PCA-guided clustering as follows:

\begin{itemize}
	\item The dimensionality of the DPD variables is first reduced using the PCA technique.
	\item The resulting data in the PCA subspace is clustered into different DPD classes using the k-means algorithm, representing varying levels of the DPD gap. The DPD status (amount of DPD measurement) can be defined from a category ordering of $M$ classes, where $c_{i}$ is the $i$-th DPD class. We arrange the DPD classes such that $c_{1}<c_{2}<\cdots<c_{M}$. 
\end{itemize}
In Section~\ref{sec:sec_4}, we show that the first PCA component (PC1) captures most variance of the DPD variables. Furthermore, we show that PC1 is sufficient for defining the clustering class of the DPD responses, i.e.,~the DPD status can be found using PC1. Accordingly, we use PC1 as the DPD index for providing one scaled measure of the DPD gap.

\begin{figure*}
	\begin{centering}
		\includegraphics[width=0.96\textwidth,trim=1cm 0cm 1cm 0cm]{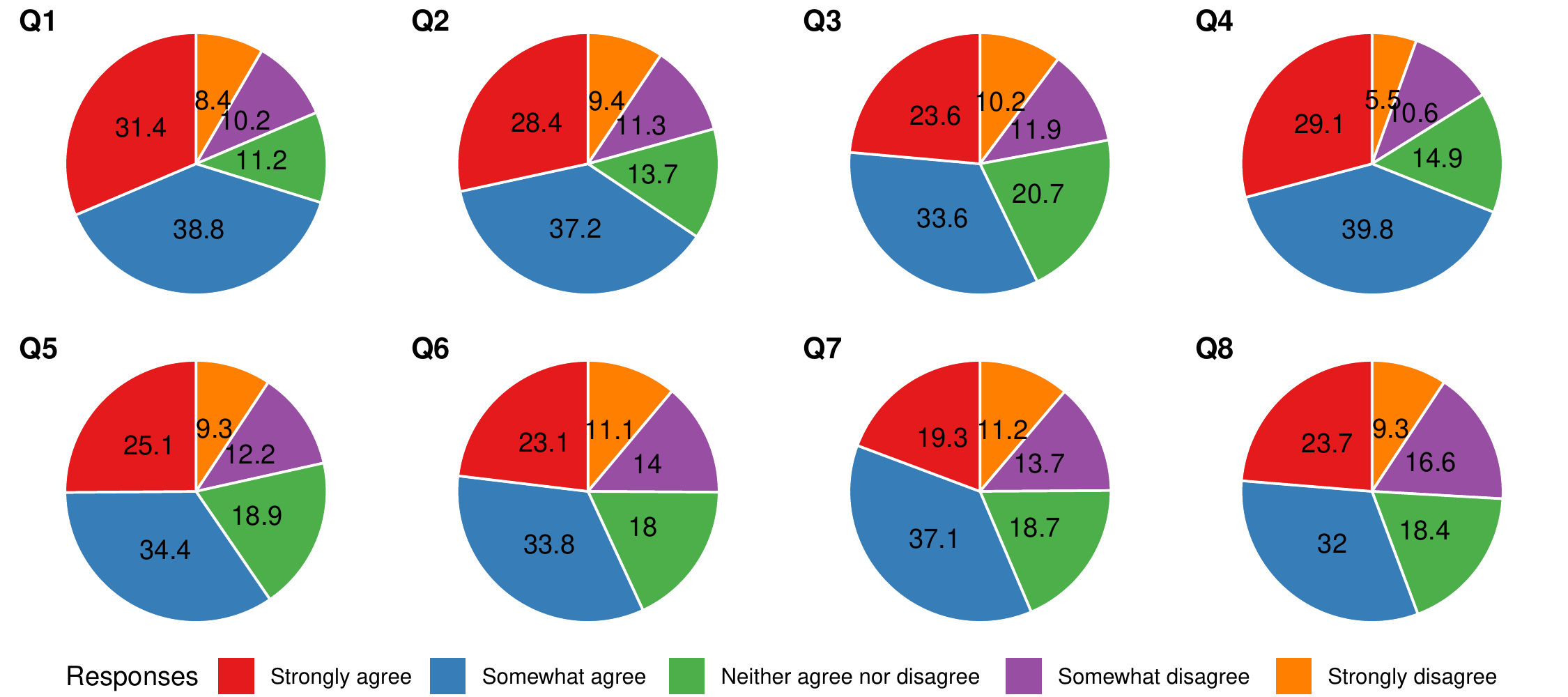}
		\par\end{centering}
	
	\caption{The distribution of responses of Questions~1-8.\label{fig:visualize_responses_pie}}
\end{figure*}

\subsection{DPD proportional odds model}\label{sec:odds_model}

We next present the DPD proportional odds model for analyzing the DPD survey data. In particular, the DPD proportional odds model enables determining the statistical correlation between a DPD status and the socio-demographic patterns of the ICT users. In addition, proportional odds regression can capture the dependency of an ordinal response on discrete or continuous variables~\cite{cox2018analysis,bilder2014analysis}.

Our objective is to define the DPD regression models which can compute the probability of each DPD class, given the socio-demographic patterns of an ICT user. Let $R$ be the DPD response random variable for the $N=776$ collected responses $R_{1},\ldots,R_{N}$. The probability of DPD class $c_i$, where $i\in[1,2,\ldots,M]$, is $\phi_{i}=\mathbf{P}(R=c_{i})$. The probability of all DPD classes is $\sum_{i=1}^{M}\phi_{i}=1$. To compute $\phi_{i}$, we must first define the cumulative probability of a DPD class. The cumulative probability of DPD class $c_{i}$ is defined as follows:
\begin{equation}
	\mathbf{P}(R\leq c_{i})=\phi_{1}+\cdots+\phi_{i},i=1,\ldots,M-1.
\end{equation}

The log-odds of the cumulative probability can be computed using the inverse of the logistic function, such that $\text{logit}\left(\mathbf{P}(R\leq c_{i})\right)=\sigma^{-1}\left(\mathbf{P}(R\leq c_{i})\right)$, where $\sigma(x)=\frac{1}{1+\exp\left(-x\right)}$. Accordingly, the log-odds of the cumulative probability can be computed as follows:
\begin{eqnarray}
	\text{\text{logit}}\left(\mathbf{P}\left(R\leq c_{i}\right)\right)& = &\log\left(\frac{\mathbf{P}(R\leq c_{i})}{\mathbf{P}(R>c_{i})}\right)\\
	& = &\log\left(\frac{\mathbf{P}(R\leq c_{i})}{1-\mathbf{P}(R\leq c_{i})}\right)\\
	& = &\frac{\phi_{1}+\cdots+\phi_{i}}{\phi_{i+1}+\cdots+\phi_{M}}.
\end{eqnarray}

In statistics, the proportional odds model can be defined as a liner combination of the explanatory variables~\cite{bilder2014analysis}. Mathematically, the log-odds of a DPD class can be computed as a liner combination of socio-demographic variables $s_{1},\ldots,s_{K}$ as follows:
\begin{equation}
	\log\left(\frac{\mathbf{P}(R\leq c_{i})}{1-\mathbf{P}(R\leq c_{i})}\right)=\eta_{i,0}+\eta_{1}s_{1}+\cdots+\eta_{K}s_{K},
\end{equation}
where $K$ is the number of socio-demographic variables collected from users. $\eta_{i,0}, \eta_{1},\ldots,\eta_{K}$ are the regression parameters. $\eta_{i,0}$~depends on the DPD class $i$.

Then, the cumulative probability $\mathbf{P}(R\leq c_{i})$ is defined as follows:
\begin{equation}
	\mathbf{P}(R\leq c_{i})=\frac{\exp\left(\eta_{i,0}+\eta_{1}s_{1}+\cdots+\eta_{K}s_{K}\right)}{1+\exp\left(\eta_{i,0}+\eta_{1}s_{1}+\cdots+\eta_{K}s_{K}\right)}.\label{eq:cumulative_probability}
\end{equation}

Using (\ref{eq:cumulative_probability}), the probability of a DPD class $c_{i}$ is computed as follows:
\begin{equation}
	\phi_{i}=\mathbf{P}\left(R=c_{i}\right)=\mathbf{P}\left(R\leq c_{i}\right)-\mathbf{P}\left(R\leq c_{i}-1\right).\label{eq:dpd_class}
\end{equation}

\section{Numerical and statistical analysis}\label{sec:sec_4}

This section presents a numerical and statistical analysis of the DPD survey. First, we provide visualizations of the responses collected using the DPD survey. Second, we present a reliability analysis of the collected responses. Third, we analyze the DPD index computation. Forth, we present a socio-demographic analysis of the DPD problem. Finally, we provide numerical results of the DPD proportional odds model.

\subsection{Visualizing and exploring DPD survey data}

Next, we present key insights of the DPD survey data using visualization charts.

\subsubsection{Distribution of responses}
Figure~\ref{fig:visualize_responses_pie} shows the distributions of responses for Questions~1-8 for all surveyed ICT users, i.e.,~all responses regardless of the socio-demographic patterns of users. Several results can be noted. First, most ICT users agree (``strongly agree'' or ``somewhat agree'') with the arguments in Questions~1-8. 70.2\%, 65.6\%, 57.2\%, 68.9\%, 59.5\%, 56.9\%, 56.4\%, and 55.7\% provided agree responses to Questions~1-8, respectively. This indicates that most ICT users are satisfied with their privacy protection. Second, a relatively high percentage of users provide a "neither agree nor disagree" response (11.2\%, 13.7\%, 20.7\%, 14.9\%, 18.9\%, 18\%, 18.7\%, and 18.4\% for Questions~1-8, respectively). This can be explained as many ICT users do not have sufficient information on their privacy protection.

\begin{figure}
	\begin{centering}
		\includegraphics[width=0.95\linewidth,trim=1cm 0.5cm 0cm 0.5cm]{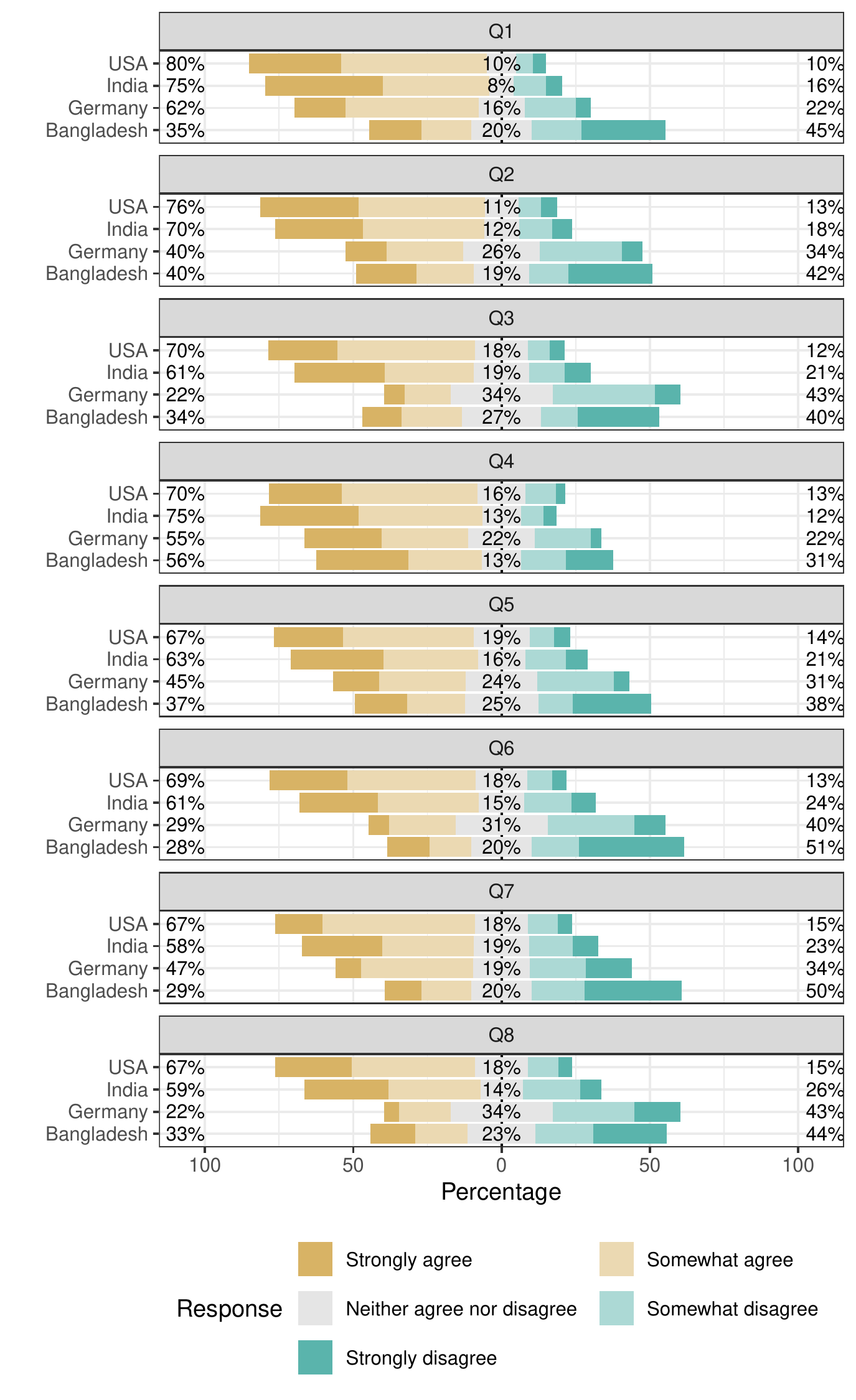}
		\par\end{centering}
	
	\caption{The distribution of responses of Questions~1-8, divided based on the countries of residency of the ICT users.\label{fig:visualize_questions_likert_q12}}
\end{figure}
Figure~\ref{fig:visualize_questions_likert_q12} shows the distribution of responses for Questions~1-8 for each country (Bangladesh, Germany, India, and the United States). The responses vary for different countries. However, there are similarities in the response percentages among India and the United States for Questions~1-8. Furthermore, the response percentages of Bangladesh and Germany look similar for Questions~2-6 and 8.

\subsubsection{Population sample and geolocation}
\begin{figure}
	\begin{centering}
		\includegraphics[width=0.95\linewidth,trim=1.5cm 0cm 1cm 1cm]{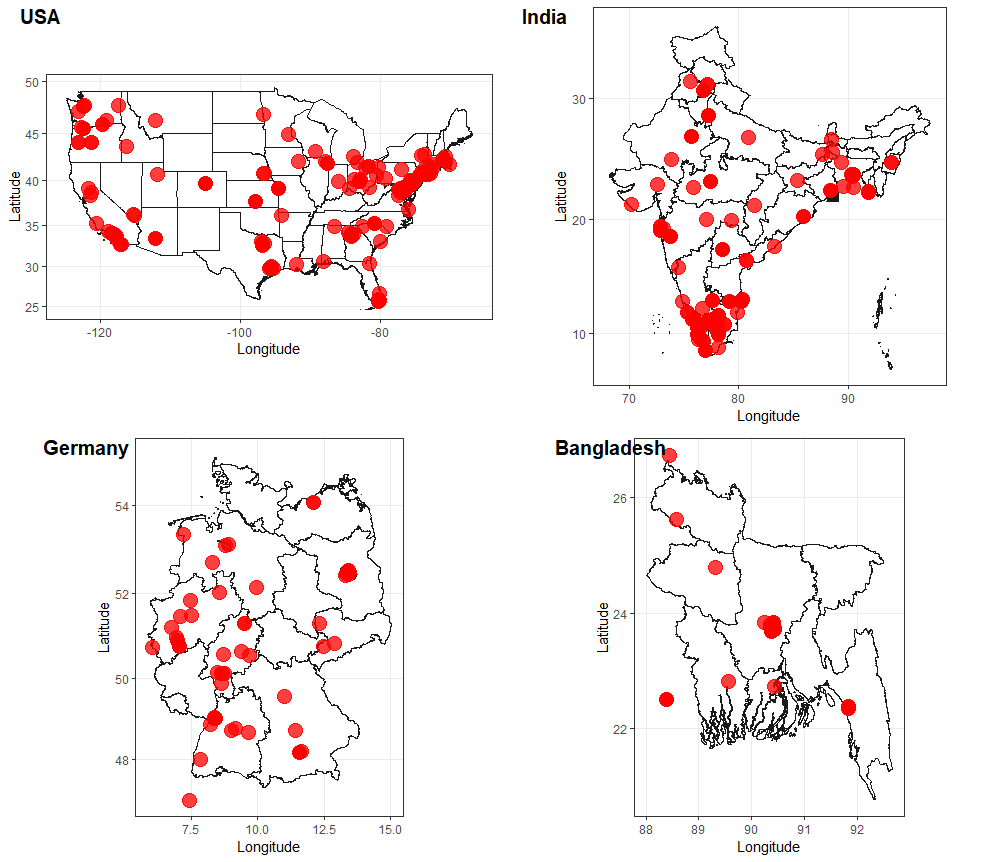}
		\par\end{centering}
	
	\caption{The geographical locations of the ICT users in Bangladesh (113~users), Germany (58~users), India (314~users), and the United States (291~users).\label{fig:visualize_responses_map}}
\end{figure}

Figure~\ref{fig:visualize_responses_map} shows the geographical locations (longitude and latitude values) of the participating ICT users in Bangladesh, Germany, India, and the United States. The numbers of collected responses are 113, 58, 314, and 291 from Bangladesh, Germany, India, and the United States, respectively. The users are located in various parts of the surveyed countries. Accordingly, the population sample of the users provides a holistic view of the countries and represents people at different geographical locations.

\subsection{Reliability of DPD survey}

We next present statistical measures for evaluating the reliability of the data collected using the online survey study. In particular, we provide the Cronbach’s $\alpha$,   McDonald’s $\omega$, and other alternative reliability measurements~\cite{zinbarg2005cronbach} of the DPD data.

\subsubsection{Cronbach’s $\alpha$}
A widely-used rule of thumb indicates that adequate internal reliability can be concluded when $\alpha$ is greater than 0.70~\cite{morera2016coefficient}. Our DPD study is reliable with internal consistency reliability of $\alpha=0.92$. This  $\alpha$ value indicates that Questions~1-8 correlate and measure the DPD problem as one construct. 

\subsubsection{Other reliability measures}
Some previous studies reported some limitations of Cronbach's $\alpha$ as a measure of reliability, e.g.,~see \cite{morera2016coefficient,mcneish2018thanks}. Therefore, we report alternative measures of reliability on the DPD data in Table~\ref{tab:reliability}. We refer the reader to~\cite{mcneish2018thanks} for an overview and mathematical definitions of these reliability measures. In summary, it can be noted that the DPD survey meets all of these reliability measures. The internal consistency of the DPD survey is concluded.

\begin{table}
	\def\arraystretch{1.3}
	\caption{Measures of internal consistency reliability.}\label{tab:reliability}
	\centering{}%
	\begin{tabular}{|c|c|}
		\hline 
		Measure & Reliability score\\
		\hline 
		\hline 
		 Cronbach's $\alpha$ & 0.92\\
		\hline 
		McDonald's $\omega$ & 0.94\\
		\hline 
		$\omega$ (hierarchical) & 0.86\\
		\hline 
		Revelle's $\omega$ (total) & 0.94\\
		\hline 
		Greatest Lower Bound (GLB) & 0.93\\
		\hline 
		Coefficient H & 0.93\\
		\hline 
		Coefficient $\alpha$ & 0.92\\
		\hline 
	\end{tabular}
\end{table}

\subsection{DPD index generation}

Next, we analyze the collected survey responses to extract the DPD index, which provides a unidimensional scale and captures the underlying communality of Questions~1-8. Subsequently, the following analysis steps are applied:
	\begin{itemize}
		\item \textit{Question (variable) analysis}: We apply factor analysis, i.e.,~aggregating the questions linearly, to capture the underlying communality of Questions~1-8. The key objective of this step is aggregating Questions~1-8 into two unobserved underlying variables called DPD factors. We will show that Questions~1-8 measure and reflect the same underlying factor.
		\item \textit{DPD index analysis}: We compute the portions of explained variance in each PCA component. Then, we analyze the influence of Questions~1-8 on the PCA components. Our analysis shows that the first principal component (PC1) captures 64.75\% of the data variation. Accordingly, PC1 is used as the DPD index, which provides one scaled measure for the DPD gap by exploring the dimensionality of Questions~1-8 in the DPD survey. 
		\item \textit{Response clustering}: It is more convenient to analyze the DPD responses using unified DPD classes, e.g.,~our DPD proportional odds model requires an ordinal variable to represent the DPD classes. Therefore, we cluster the responses into DPD classes of 4 levels using k-means clustering. Furthermore, we show that PC1 is sufficient for defining the DPD class of any response.
	\end{itemize}
	
\subsubsection{Question (variable) analysis}

\begin{figure}
	\begin{centering}
		\includegraphics[width=0.95\linewidth,trim=1cm 1cm 1cm 1cm]{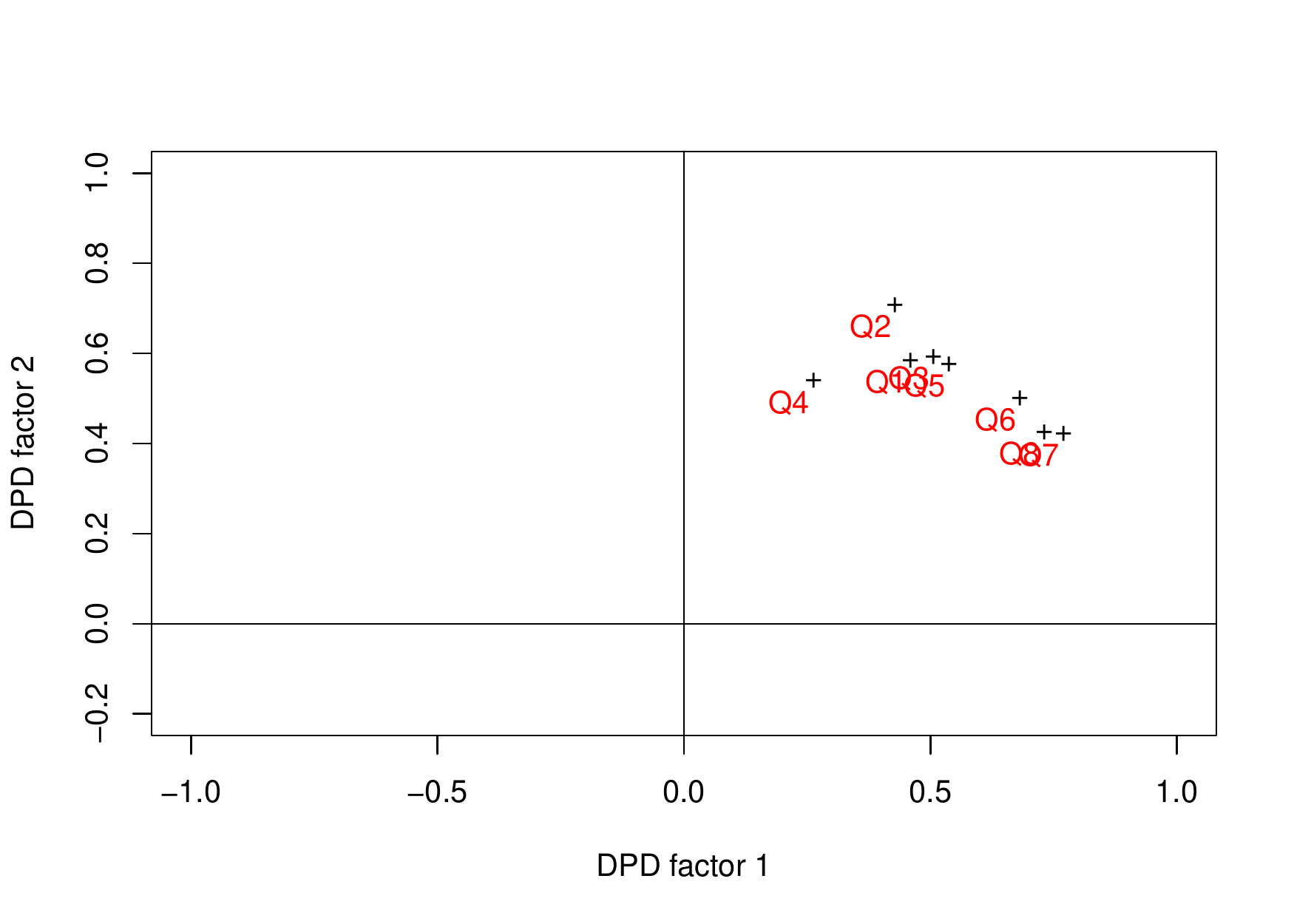}
		\par\end{centering}
	\caption{Factor analysis of Questions~1-8.\label{fig:evaluate_factor_analysis}}
\end{figure}

Figure~\ref{fig:evaluate_factor_analysis} shows the factor analysis of Question~1-8. The main objective of the factor analysis is capturing the underlying communality of the questions by using a linear combination of factors~\cite{hartmann2018learning}, i.e.,~factor analysis enables understanding the underlying DPD concept by aggregating the questions. The contributions of questions to the factor is shown as points in Figure~\ref{fig:evaluate_factor_analysis}. It can be noted that Questions~1-8 are located near each other. This shows that Questions~1-8 capture the underlying DPD gap as a single construct.

\subsubsection{DPD index analysis}
\begin{figure}
	\begin{centering}
		\includegraphics[width=0.95\linewidth,trim=1.5cm 1cm 0cm 0.5cm]{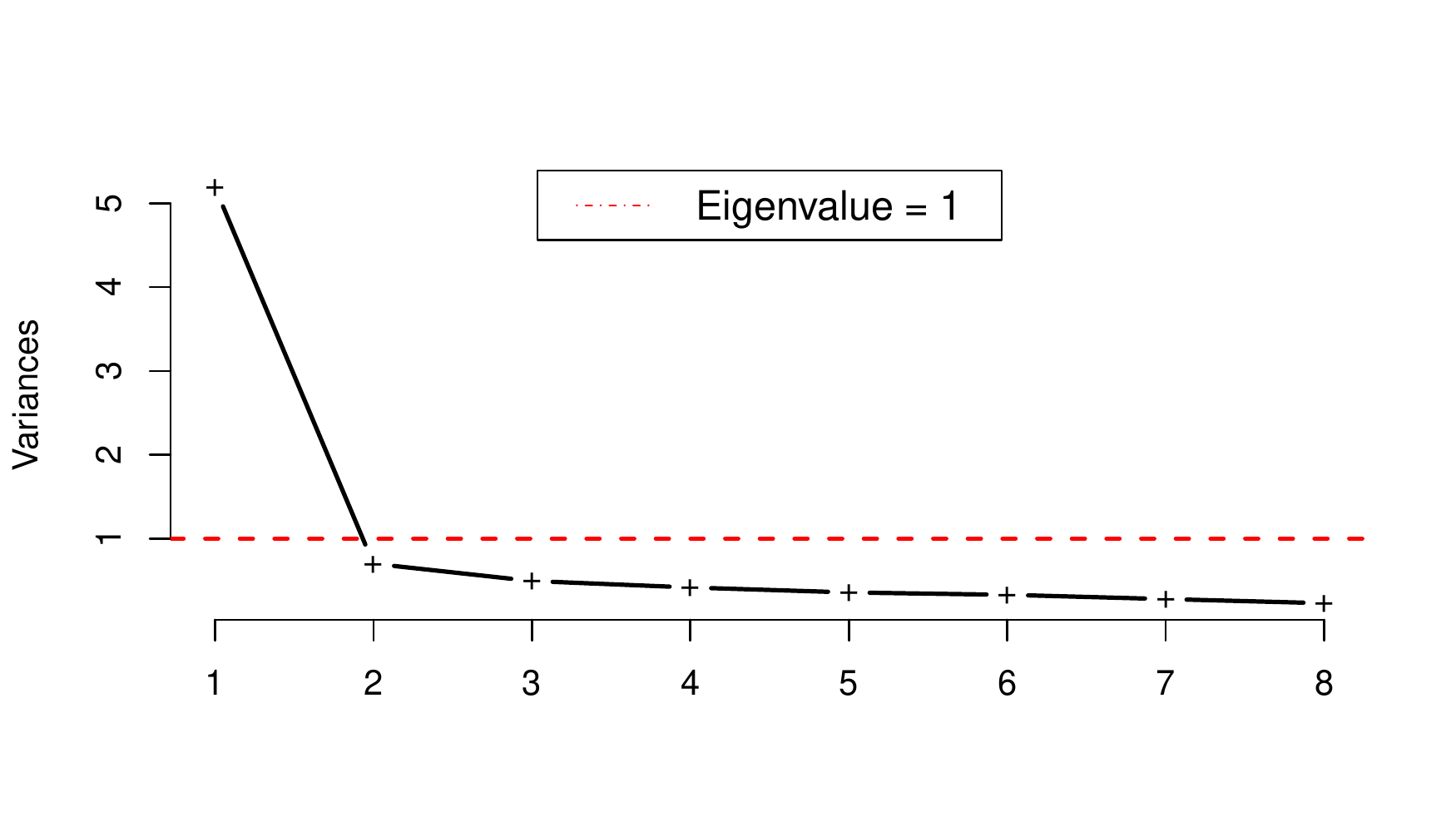}
		\par\end{centering}
	\caption{Variance captured by each PCA component.\label{fig:compute_privacy_index_1}}
\end{figure}

Figure~\ref{fig:compute_privacy_index_1} shows the portions of explained variance and eigenvalues in each of the DPD components using PCA. The x-axis shows the number of DPD principal components, and the y-axis shows the variance explained by each principal component. For example, the variance explained by the first principal component is 5.18. It can be noted that all PCA components, except the first one, have eigenvalues of less than 1. Given that there is only one PCA component with an eigenvalue of 1.0 or higher, it can be concluded that the questions form a unidimensional scale, and they are internally consistent.

\begin{figure}
	\begin{centering}
		\includegraphics[width=0.92\linewidth,trim=0.5cm 0cm 0.5cm 0cm]{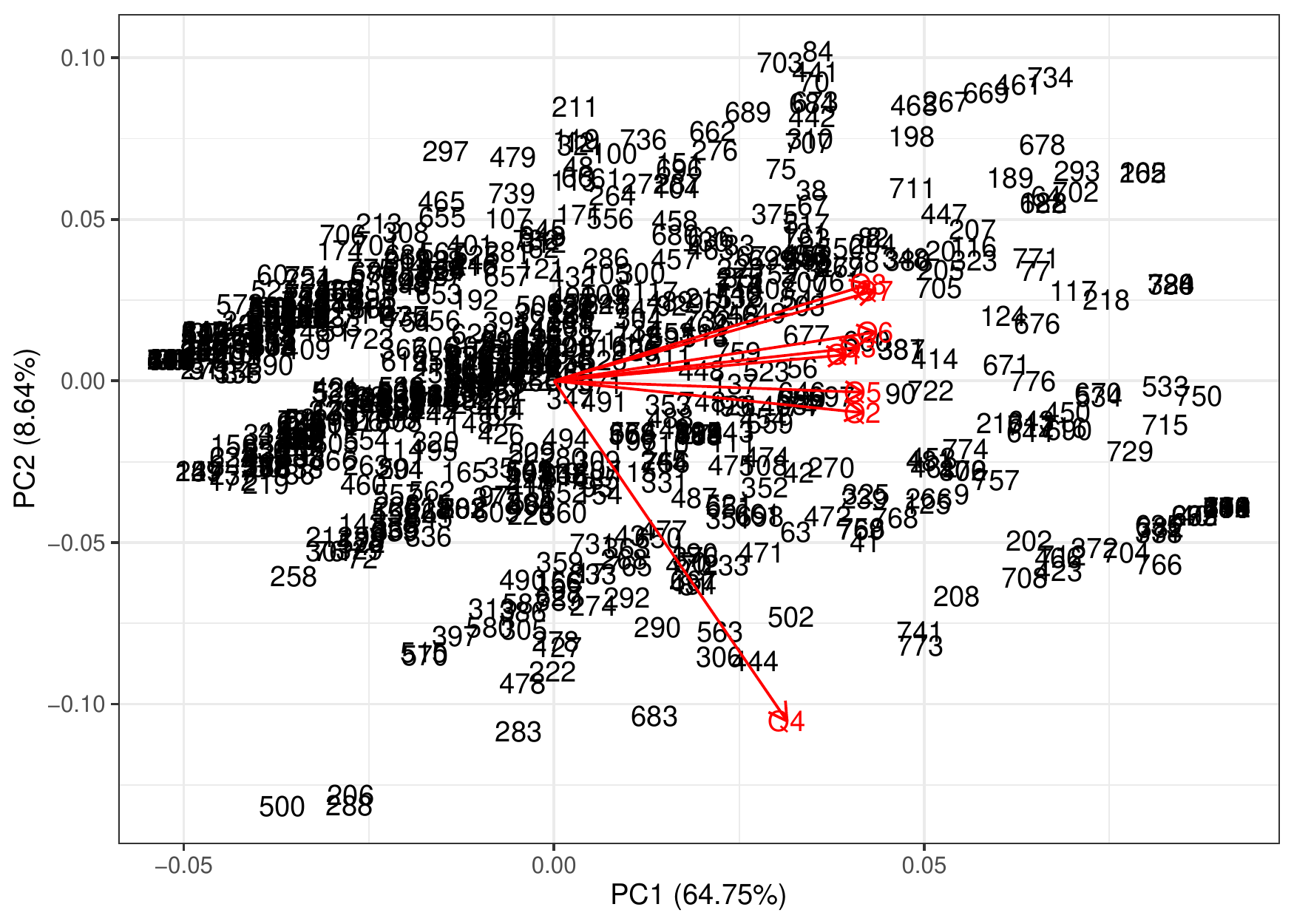}
		\par\end{centering}
	\caption{Exploratory graph showing information on both response samples and question variables of the DPD survey data.\label{fig:compute_privacy_index_2}}
\end{figure}

Figure~\ref{fig:compute_privacy_index_2} shows both the PCA scores (data points) and loading values (red vectors pinned from the origin of PC1 and PC2). Several results can be observed. First, the first principal component (PC1) captures 64.75\% variation of the data, i.e.,~PC1 explains the majority of the variance in the DPD variables. In comparison, the second principal component (PC2) captures only 8.64\% of the data variation. Second, a loading vector reflects the degree to which each question influences the computation of PCA components. For example, Question 4 has a strong influence when computing PC2. All questions have a strong influence on the computation of PC1. Third, the score points show the projections of the responses into the PCA subspace. The DPD gap, i.e.,~various levels of privacy protection for users, can be observed.

\subsubsection{Response clustering}
\begin{figure}
	\begin{centering}
		\includegraphics[width=0.95\linewidth,trim=1cm 0cm 1cm 0cm]{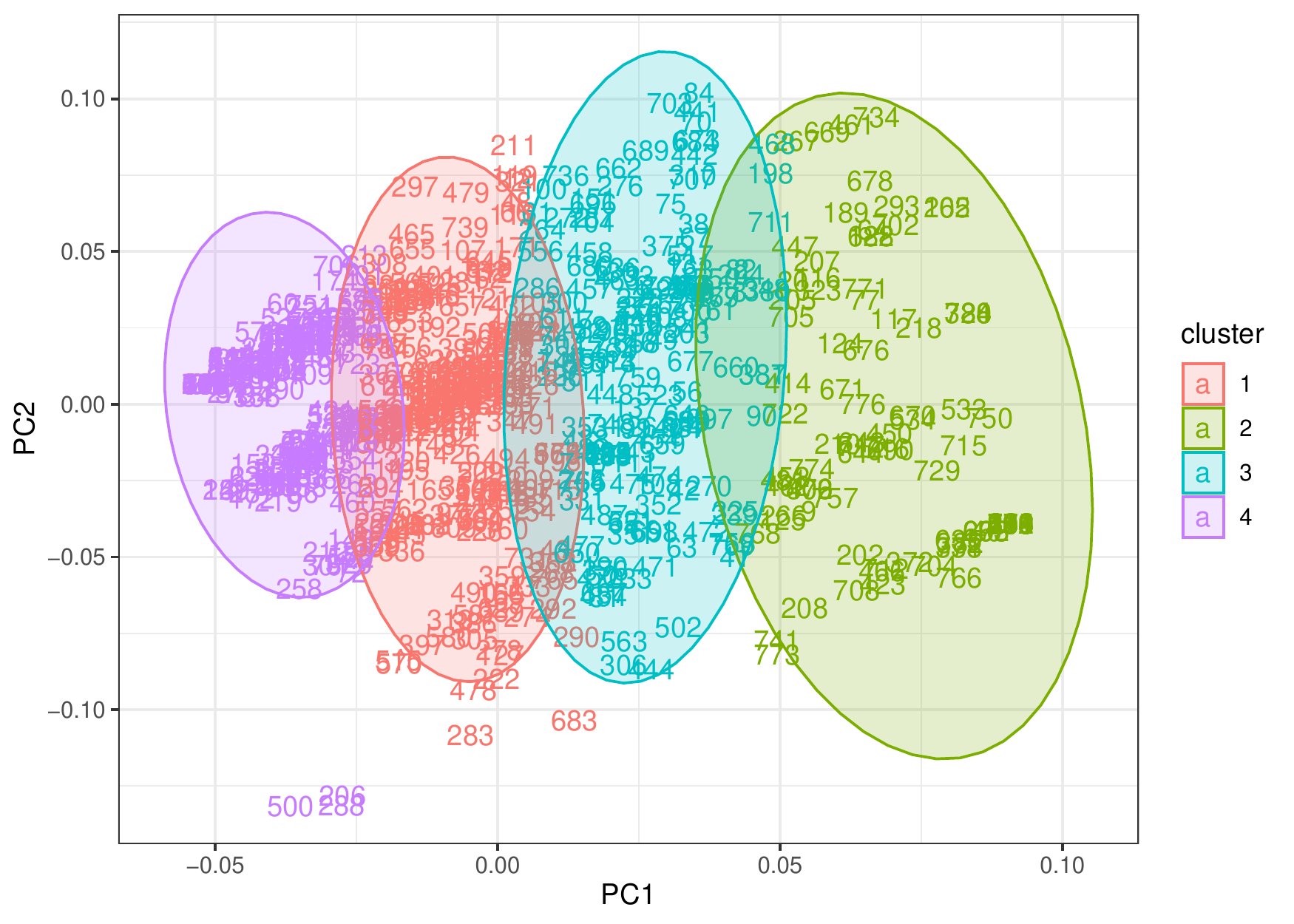}
		\par\end{centering}
	\caption{Clustering the responses into 4 DPD classes ($M=4$).\label{fig:compute_privacy_index_3}}
\end{figure}

Figure~\ref{fig:compute_privacy_index_3} shows the clustering of the observable variables (Questions~1-8) into 4 DPD classes. Each DPD class represents a different level of the DPD gap. It can be noted that the DPD classes are defined based on the first principal component (PC1) only, i.e., the clustering ellipses are vertical. The major axes of the ellipses are parallel to PC2, and PC2 is not used to define the DPD status. Therefore, PC1 is used to represent the DPD index in the next socio-demographic analysis.

\subsection{Socio-demographic analysis of DPD}

Next, we provide an in-depth experimental discussion of the socio-demographic analysis in the DPD problem.

\begin{figure*}
	\begin{centering}
		\includegraphics[width=0.95\textwidth,trim=0cm 1cm 0cm 0.5cm]{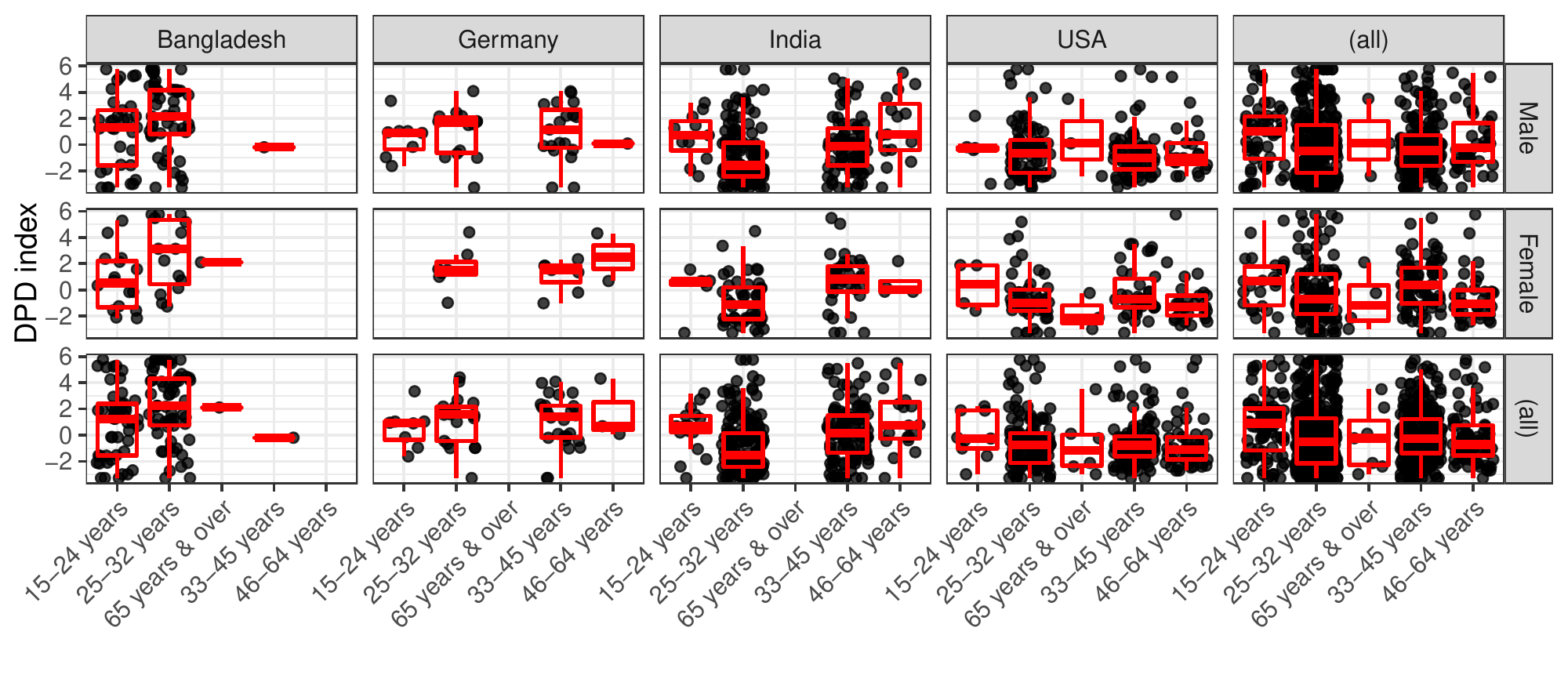}
		\par\end{centering}
	\caption{The DPD index based on the age groups of the ICT users.\label{fig:visualize_facet_grid_q13}}
\end{figure*}

Figure~\ref{fig:visualize_facet_grid_q13} shows the distribution of responses based on the ages, genders, and countries of residency of the ICT users. 67.7\% and 32.3\% of the surveyed ICT users are males and females, respectively. The percents of collected responses are 14.6\%, 7.5\%, 40.5\%, and 37.5\% from Bangladesh, Germany, India, and the United States, respectively. Several results can be drawn. First, the median DPD index is significantly high in Bangladesh and Germany, compared to India and the United States (see the third row). This reflects significant concerns on privacy protection among the ICT users in Bangladesh and Germany. Second, there is no significant difference in the median DPD index among all ICT users based on their genders (see the first and second rows). Third, there are high variations among Bangladesh's ICT users (see the median, first quartile, and third quartile in the first column).

\begin{figure*}
	\begin{centering}
		\includegraphics[width=0.95\textwidth,trim=0cm 1cm 0cm 0.5cm]{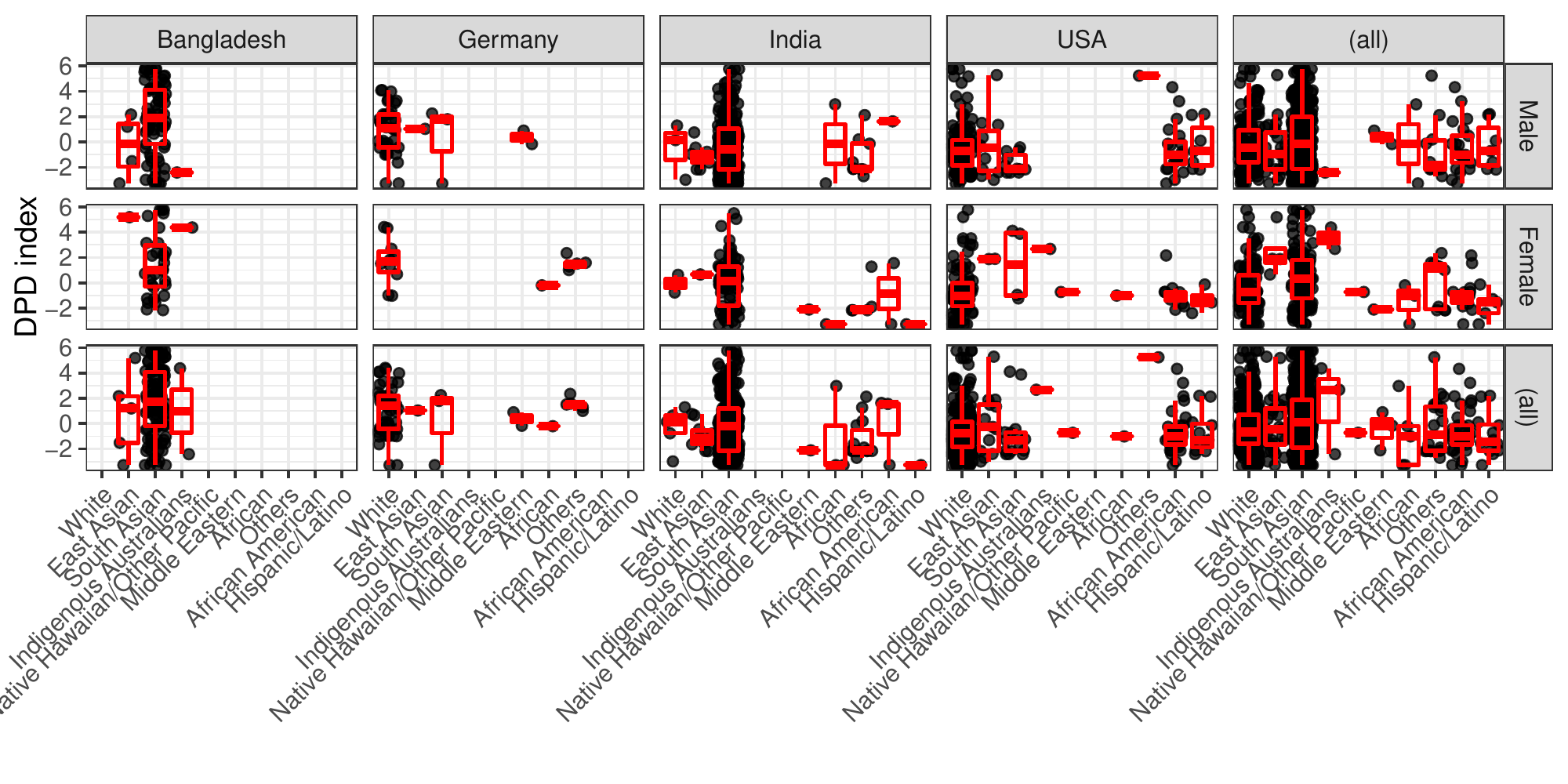}
		\par\end{centering}
	\caption{The DPD index based on the ethnic groups of the ICT users.\label{fig:visualize_facet_grid_q14}}
\end{figure*}

\begin{figure*}
	\begin{centering}
		\includegraphics[width=0.95\textwidth,trim=0cm 1cm 0cm 0.5cm]{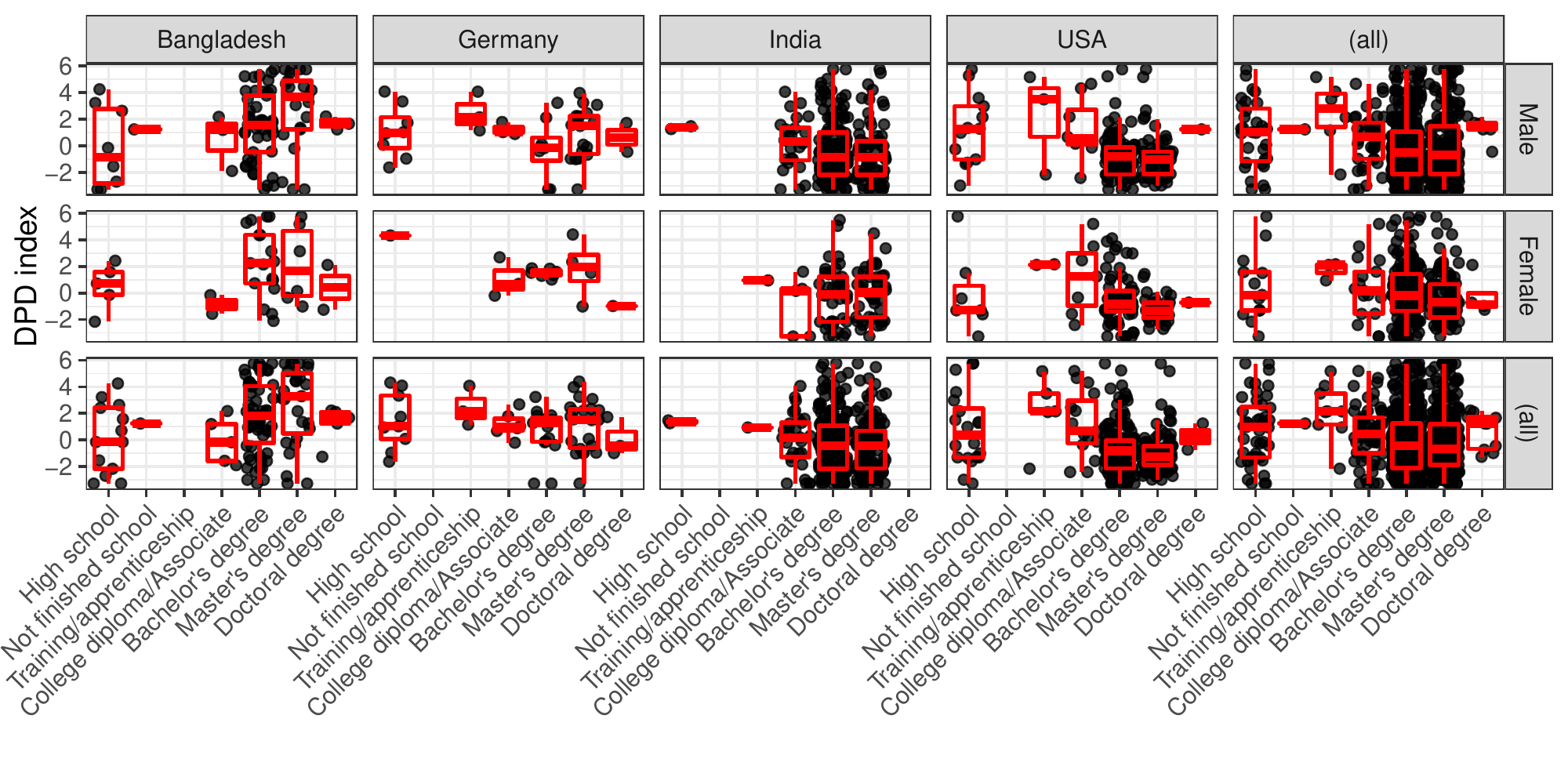}
		\par\end{centering}
	\caption{The DPD index based on the highest levels of education of the ICT users.\label{fig:visualize_facet_grid_q16}}
\end{figure*}

\begin{figure*}
	\begin{centering}
		\includegraphics[width=0.95\textwidth,trim=0cm 1cm 0cm 0.5cm]{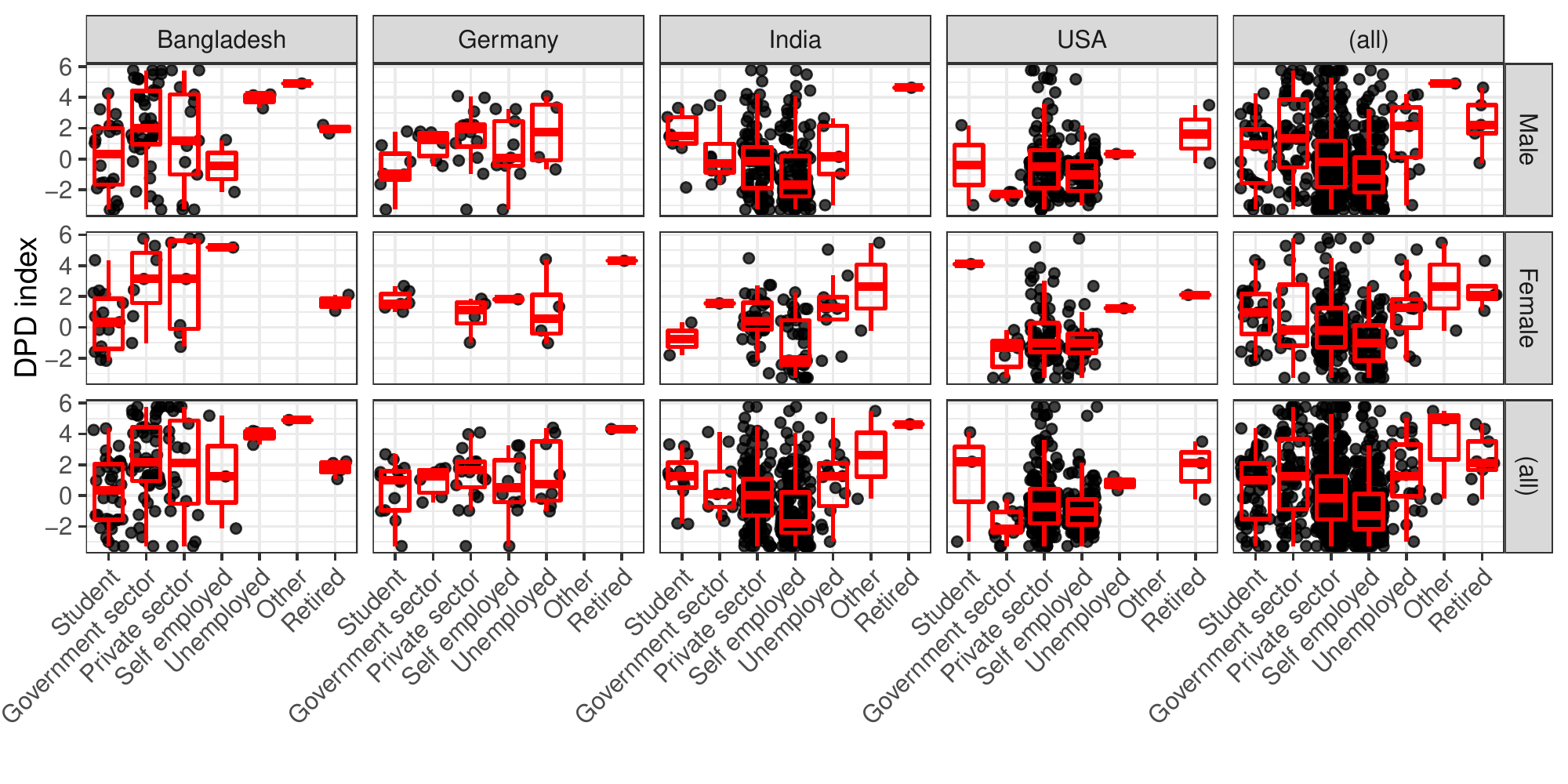}
		\par\end{centering}
	\caption{The DPD index based on the occupations of the ICT users.\label{fig:visualize_facet_grid_q17}}
\end{figure*}

Figure~\ref{fig:visualize_facet_grid_q14} shows the distribution of responses based on their ethnic backgrounds, genders, and countries of residency. The surveyed ICT users come from various ethnic backgrounds, including South Asians (51.7\%), Whites (34.4\%), African Americans (4.5\%), and East Asians (3.7\%). Figure~\ref{fig:visualize_facet_grid_q16} shows the distribution of responses based on their highest levels of education, genders, and countries of residency. The surveyed ICT user have various levels of education, such as bachelor's degrees (51.8\%), master's degrees (32.9\%), college diplomas (7.1\%), and high school graduates (5.7\%). Figure~\ref{fig:visualize_facet_grid_q17} shows the distribution of responses based on their occupations, genders, and countries of residency. The surveyed ICT users work in all occupation sectors, including the private sector (42.4\%), self-employment (34.9\%), and the government sector (9.7\%).

\subsection{DPD proportional odds model}

\begin{figure*}
	\begin{centering}
		\includegraphics[width=0.65\textheight,trim=1cm 0cm 1cm 0cm]{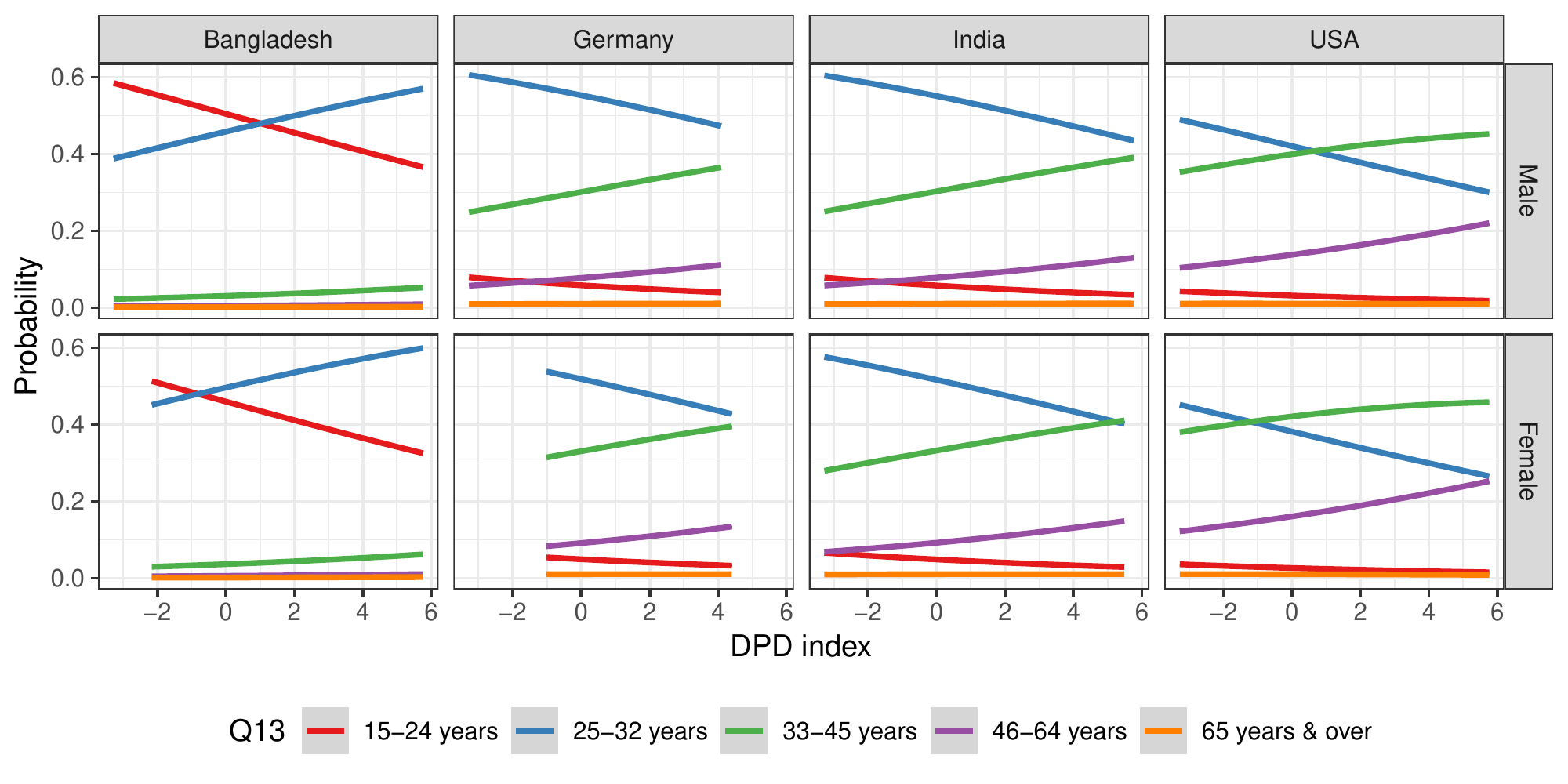}
		\par\end{centering}
	\caption{Ordinal regression of the DPD index using the  socio-demographic patterns of the ICT users (country of origin, gender, and age group).\label{fig:ordinal_regression_q13}}
\end{figure*}

Finally, we analyze the responses based on the DPD proportional odds model proposed in Section~\ref{sec:odds_model}. Figure~\ref{fig:ordinal_regression_q13} shows the predicted probability of reporting a DPD index based on the age group of an ICT user. The likelihood of providing high DPD index increases rather dramatically with age. This result reflects more concerns among young users (15-32 years) on their privacy protection than senior users (33 years and over). The only exception is among the 25-32 years old users in Bangladesh, which tend to indicate less concern about their digital privacy.

We also applied the DPD proportional odds model on the ethnic backgrounds, occupations, and highest education levels of the ICT users. We find that the ethnic backgrounds, occupations, and highest education levels have a minimal statistical impact on the DPD gap perceived by the ICT users.

\section{Roadmap discussion}\label{sec:roadmap}

The DPD problem is a recent form of the digital divide. The DPD gap results in severe financial, social, and physiological difficulties for the exposed users. It can take several years for the exposed victims to recover from a digital privacy breach. The stolen data is generally used in criminal activities~\cite{aimeur2011ultimate}. Therefore, all organizations and institutions must invest in protecting the privacy of their ICT users by applying state-of-the-art privacy tools~\cite{sahi2017privacy,xu2014information,nguyen2021security,gupta2020security}.

We recommend the following roadmap for addressing the DPD problem.

\subsubsection{Digital privacy regulations must be introduced in all countries}
Legislative bodies can play the most significant role in closing the DPD gap. Therefore, we recommend introducing data protection laws comparable to the GDPR by all countries. Furthermore, privacy protection must be enabled as the default preference for all ICT users, regardless of their socio-demographic patterns or country of residency.

\subsubsection{Researchers and educators are obligated to increase the awareness of the DPD problem}

Teaching curriculums must include significant components covering digital privacy. In addition, the media can provide a valuable medium for reaching the general audience. We must increase the awareness of digital privacy as a fundamental human right.

The majority of the research in cybersecurity focuses on data and network security (preventing unauthorized access to data by third parties). In contrast, digital privacy (regulating how to collect, process, and share data) does not receive equal attention within the research community. Thus, more awareness within the research community will encourage future works on digital privacy.

\section{Conclusion and Future work}\label{sec:conclusions}
In this paper, we have presented a survey study for understanding the DPD problem. We used crowdsourcing task assignments to collect responses from 776~ICT users on the DPD problem. The DPD survey is shown to meet the internal consistency reliability, including  Cronbach’s $\alpha$ of $0.92$ and McDonald’s $\omega$ of $0.94$. Furthermore, the DPD index is shown to capture the underlying DPD construct using the PCA-guided clustering method. Finally, we have explored the statistical relationship between the DPD problem and the socio-demographic patterns of the ICT users using the DPD proportional odds model.

Future studies and relevant industries can pursue several important directions based on the results of this paper.
\begin{itemize}
	\item There is an urgent need to build third-party privacy impact analysis and compliance checks in ICT systems. For example, privacy star ratings can be issued to ICT systems based on their compliance with the privacy provisions.
	\item Open-source privacy tools and analytics opt-out extensions are still underdeveloped. At the same time, there is a necessity to study the privacy protection achieved through existing privacy tools, such as the Google Analytics opt-out browser add-on.
	\item Future research is warranted to understand privacy breaches' social, economic, and cultural impacts on individuals and institutions.
\end{itemize}

\bibliographystyle{IEEEtran}
\bibliography{reference}
\section*{\textbf{Biographies}}
\begin{IEEEbiography}[{\includegraphics[width=1in,height=1.25in,clip,keepaspectratio]{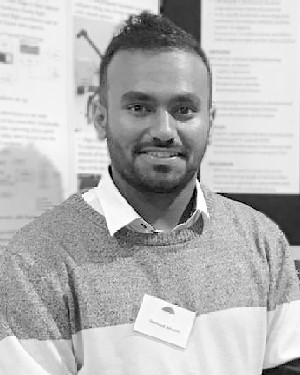}}]{Hamoud Alhazmi} (S'21)
is working as a Research Assistant at the University of Canberra, ACT, Australia. He graduated with a Master's degree in Cybersecurity and B.Eng. degree in Network \& Software Engineering with (first-class honors) from the University of Canberra in 2021 and 2020, respectively. His current research interests are in computer vision, machine learning, and cybersecurity. He worked as an Assistant IT Manager during his studies in Canberra.
\end{IEEEbiography}

\begin{IEEEbiography}[{\includegraphics[width=1in,height=1.25in,clip,keepaspectratio]{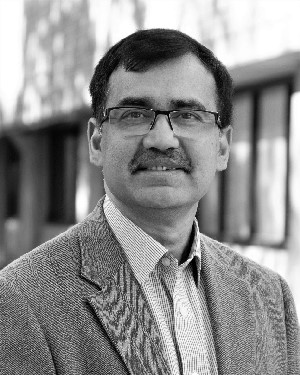}}]{Ahmed Imran}
(ahmed.imran@canberra.edu.au) is an Information Systems researcher at the University of Canberra with special interests in the strategic use of IT, eGovernment, and socio-cultural impacts of ICT. His vast experience as an IT manager as well as his work in developing countries became invaluable for research and in understanding and providing a rich insight into the socio-cultural context through multiple lenses, resulting in interdisciplinary research opportunities. His research has proven to bring real-world applications to the table, something that cemented its importance and relevance in the eyes of the research community. This recognition was further reflected through the award of the prestigious Australian National University Vice Chancellor's award in 2010, followed by numerous invitations to international and national forums/universities.
\end{IEEEbiography}

\begin{IEEEbiography}[{\includegraphics[width=1in,height=1.25in,clip,keepaspectratio]{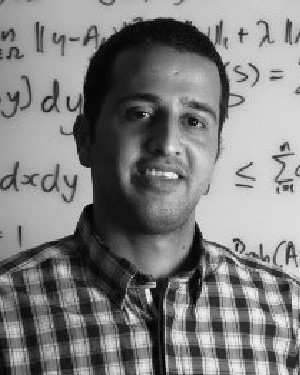}}]{Mohammad Abu Alsheikh} (S'14--M'17)
is an Associate Professor and ARC DECRA Fellow at the University of Canberra (UC), ACT, Australia. He designs and creates novel privacy-preserving Internet of things systems that leverage both machine learning and convex optimization with applications in people-centric sensing, human activity recognition, and smart cities. Previously, he was a Postdoctoral Researcher at the Massachusetts Institute of Technology (MIT), USA. His doctoral research at Nanyang Technological University (NTU), Singapore, focused on optimizing wireless sensor networks' data collection. After graduating with a B.Eng. degree in computer systems from Birzeit University, Palestine, he worked as a software engineer at a digital advertising start-up and Cisco.
\end{IEEEbiography}

\vfill

\end{document}